# SIMULATION AND OPTIMIZATION OF CONTAINER TERMINAL OPERATIONS: A CASE STUDY

**GAMAL ABD EL-NASSER A. SAID[1], ABEER M. MAHMOUD [2], EL-SAYED M. EL-HORBATY[3]**

Computer science Department, Faculty of Computer & Information Sciences, Ain Shams University, Egypt.

**ABSTRACT**

Container terminals are facing a set of interrelated problems. Container handling problems at container terminals are NP –hard problems. The docking time of container ships at the port must be optimized. In this paper we have built a simulation model that integrates all the activities of a container terminal. The proposed approach is applied on a real case study data of container terminal at El-Dekheilla port. The results show that the proposed approach reduced the ship turnaround time in port where 51% reduction in ship service time (loading/unloading) in port is achieved.

## INTRODUCTION

With the rapid trade globalization, the marine transportation is getting more and more popular. Large numbers of cargos are moved in containers through ports. Therefore, effective and efficient management of port container terminals is quite important in marine transportation development [5]. Thousand of containers are handled in a container terminal everyday by different types of material handling equipment. Managing activities of such high intensity level in a container terminal is a challenging task. In a container terminal, the allocation of resources is typically triggered by time, and all resources have to be considered simultaneously in the terminal's resource allocation process [9]. In container terminals, different types of material handling equipment are used; quay cranes, yard cranes and trucks. Such equipments are used in different parts of a container terminal to transfer containers from one location to another. The docking time of container ships at the port must be as small as possible. This means that container handling process has to be completed in a short time, with a minimum use of different expensive equipments. Therefore many researchers try to find or improve different methods to solve these problems with high quality and in less time. When a ship arrives at the port, multiple quay cranes are assigned for discharging the import containers from ships to trucks. Trucks transport discharged containers to storage yards. Yard Cranes are assigned for storing containers at storage yards for a certain period, there are different types as well as different sizes of containers; the type and the size of container have an effect on allocation of containers to the storage blocks. Import containers that are handling through the discharging and receiving processes, while export containers that are handling through the loading and delivering processes [21]. In a container terminal (Figure. 1), vessels arrive based on some predetermined schedules. Quay cranes unload containers from vessels to yard trucks or load containers from yard trucks to vessels. Yard trucks travel between quay cranes and yard cranes to transport containers. Yard cranes in the yard and handle containers between yard trucks and container stacks [30].

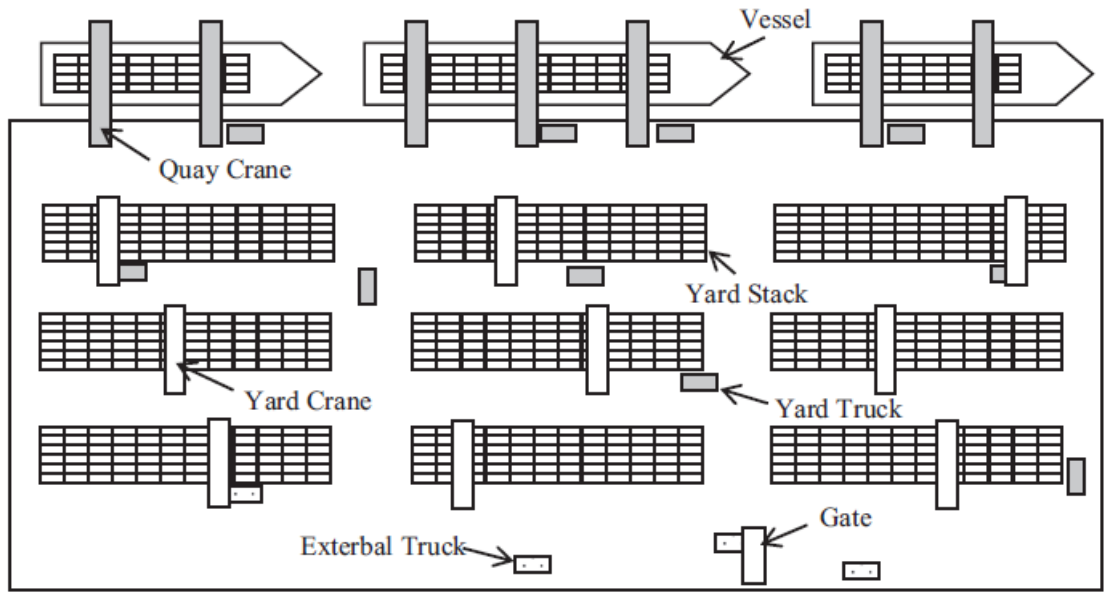

Figure. 1. A typical container terminal and its equipment (Zhuo Sun et al., 2012)

There are many different decisions involved in operating container terminals and all these decisions affect each other. For example, decisions about the storage of containers in the yard directly affect the workloads of the yard cranes in the blocks and the traveling distances of the Internal Trucks (ITs) and indirectly affect the efficiency of Quay cranes (QCs). All these decisions are also related to the berth allocation of vessels [28]. Container handling problems at container terminals are NP-hard, stochastic, nonlinear, and combinatorial optimization problems, thus requiring the application of heuristic algorithms to reach solutions [7,13,16]. Met-heuristics algorithms have been used to solve optimization problems, Among all of the heuristic algorithms such as : genetic Algorithm, tabu Search, and simulated annealing, genetic algorithms (GAs) are in wide application because of their ability to locate the optimal solution in the global solution space [3,5,14]. In general terms, container terminal can be described as open systems of material flow with two external interfaces which are the quayside with loading and unloading of ships, and the landside where containers are loaded and unloaded on/off trucks. Containers are stored in stacks thus facilitating the decoupling of quayside and landside operation [23]. Several NP-hard combinatorial optimization problems, such as the traveling salesman problem and yard management of container terminals can be modeled as Quadratic Assignment Problem QAP [15, 27].

Discrete event simulation has long been a useful tool for evaluating the performance of complex systems. Simulation is a tool to evaluate the performance of a system, under different configurations of interest and over long periods of real time. In discrete event simulation the central assumption of the system changes instantaneously in response to certain discrete events [8, 21]. Container handling problems at container terminal are too complex to be modeled analytically; discrete event simulation has been a useful tool for evaluating the performance of systems. However, simulation can only evaluate a given design, not providing optimization of such systems [10]. Discrete-event simulation exploited to support container terminal decisions in a complex and stochastic environment. Simulation-based approaches have been widely used to model various planning problems arising in container terminals [11]. A simulation optimization model for scheduling loading operations in container terminals is developed to find good container loading sequences which are improved by a genetic algorithm through an evaluation process by simulation model to evaluate objective function of a given scheduling scheme [10].

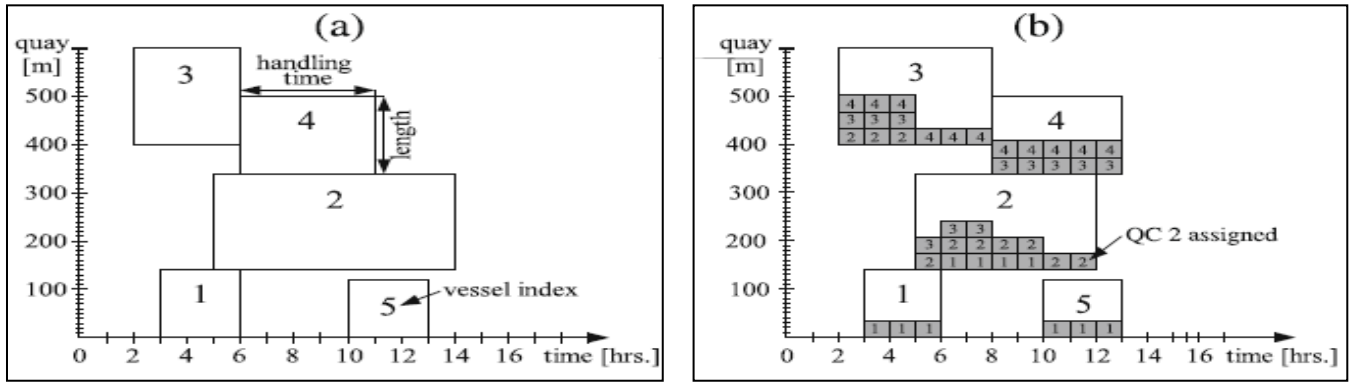

Figure.2. Space-time representation of a berth plan (a), assignment of cranes to vessels (b). (Bierwirth and Meisel, 2010)

The port operator aims at maximizing port efficiency, which includes minimizing the vessel turnaround time, maximizing the quay crane productivity, and maximizing the throughput of container terminals [30]. All vessels must be moored within the boundaries of the quay. They are not allowed to occupy the same quay space at a time. The problem is to assign a berthing position and a service time to each vessel, such that a given objective function is optimized. An example for the graphical representation of a berth plan with five vessels is shown in Figuer.2a. The cranes are supposed to be lined up alongside the quay. They can be moved to every vessel but they are not able to pass each other. The problem is to assign cranes to vessels such that all required transshipments of containers can be fulfilled, see Figure.2b [4]. In container terminals, many combinatorial related problems and the solution of one of the problems may affect to the solution of other related problems. For instance, the berth allocation problem can affect to the crane assignment problem and both could also affect to the Container Stacking Problem [22].

The rest of this paper is organized as follows: Section II presents the literature review for solving container terminal problems. Section III discusses the methodology and framework developed for simulation container terminal operation. Experimental results are given in section IV. Conclusions and future work are given in section V.

**LITERATURE REVIEW**

A container terminal works under multiple operational objectives. Container terminal operations comprise of an intricate set of container handling processes. Many approaches have been developed to solve container terminal problems. Cordeau et al. (2007) study the Service Allocation Problem, the objective is the minimization of container handling operations in the yard and it is formulated as a Quadratic Assignment Problem [15]. Silva et al. (2008) presents a comparative study on Genetic algorithm (GA) and Ant Colony Optimization (ACO), in order to decide which of the algorithms should be used on the optimization and rescheduling of a logistic system. The study is based on two different analyses: a literature survey, where the performances of GA and ACO are compared for different well-known benchmark problems, and on a detailed comparison for a logistic optimization problem. Both analyses led to the same general conclusions: GA and ACO have good and similar performances for different instances of different optimization problems, but the GA have in general a lower computational burden. However, the extra-time required by the ACO is used to record information about the optimization procedure that can be used to reschedule the logistic system in dynamic environments [26]. Jinxin et al. (2008) considered the problem of scheduling of trucks in a container terminal to minimize makespan. An integer programming model for truck scheduling and storage allocation problem is formulated. [16]. Huang et al. (2008) presents a simulation model that can be used to simulate the container terminal operations for the purpose of terminal design, capacity planning and operations planning, it was found to be effective in replicating real-world operations as well as in evaluating the handling capacities [18].

K.L. Mak and D. Sun (2009) presents a new hybrid optimization algorithm combining the techniques of genetic algorithm and tabu search method to solve the problem of scheduling yard cranes to perform a given set of loading and unloading jobs in a yard zone [1]. Mohammad. B. et al. (2009) solved an extended Storage Space Allocation Problem (SSAP) in a container terminal by GA. The objective of the SSAP developed is to minimize the time of storage and retrieval time of containers [24]. Zeng and Yang (2009) presents a framework of simulation optimization. Skinner et al. (2012) presents a modified mathematical model incorporates Quay cranes (QCs) related operations. GA-based approach was presented to solve the job scheduling problem. The proposed approach has been fully implemented on a trial basis in the scheduling system and it effectively improves the performance of the container terminal [6]. Simulation modeling techniques are being applied to a wide range of container terminal planning processes and operational analysis of container handling systems. Legat et al. (2012) present queuing-based representation of the housekeeping process in a real container terminal and solve it by discrete-event simulation to i) assess the efficiency of the housekeeping operations under unforeseen events or process disturbances and ii) estimate the related productivity and waiting phenomena which, in turn, affect the vessel turn-around time [21].

Sriphrabu (2013) proposes a developed simulation model for stacking containers in a container terminal through developing and applying a genetic algorithm (GA) for containers location assignment with minimized total lifting time and increased service efficiency of the container terminals [22]. Kotachi et al. (2013) presents a generic discrete-event simulation that models port operations with different resource types including rubber tyred gantry cranes (RTG), quay cranes, trucks, arriving and departing ships. The analysis study various scenarios motivated by changes in different inputs to measure their impact on the outputs that include throughput, resource utilization and waiting times [25].

Diabat et al. (2014) presents a Genetic Algorithm (GA) to solve Quay Crane Assignment problem, the performance of the GA in terms of solution quality was compared to the exact solution. Computational results indicate that the GA produces solutions with a small gap from the optimal solution. The most important characteristic of the developed model is the integration of the assignment and scheduling problem for quay cranes, which yields better results than solving these problems independently [2]. Abd El-Nasser et al. (2014) presents a model using computer-based modeling that can be used to simulate the operations and analyze the performance of container terminal for improving the efficiency of container terminal [29].

**METHODOLOGY AND FRAMEWORK**

This paper focuses on the urgent need of highlights that there is a great need of improving the efficiency of container terminal operation. Measuring efficiency is extremely important to terminal operation and helps terminal operators to ensure that they are getting optimal use from their equipment. A key measure of efficiency of terminal service is the ship turnaround time. The objective of operation optimization is minimizing the ship turnaround time. Therefore, it is very important to make ship turnaround time as short as possible. A computer simulation is one such area that can really help in improving the performance of a container terminal without extending the number of resources but using the same available equipments with proper planning. In this study, we have built a simulation model that can be used to simulate container terminal operation. The process of ship handling and berth allocations was analyzed. The proposed approach is applied on a real case study data from Container Terminal at El-Dekheilla port.

**Case Study: Container Terminal at El- Dekheilla port.**

Our proposed approach is applied on a container terminal at El- Dekheila port. The scope of our model is as follows:

| | |
|---|---|
| Container quay length | 1040 meter |
| Water Depth | 12-14 meter |
| Terminal Area | 406000 m2 |
| Number of Quay cranes | 8 |
| Number of Yard Cranes | 12 |
| Number of Trucks | 45 |
| Number of Heavy Top Lift Truck | 16 |
| Number of Empty Handler Side Spreader | 5 |
| Number of Mobile Crane Capacity 100Ton | 1 |

**Data collection and preparation**

The data used in this study was obtained from El-Dekheilla container terminal. The daily "log sheets" of the planning department at El-Dekheilla container terminal include the detailed information for each ship arrived to the port; every sheet contains the following data: ship Name, ship length, agent name, type; size and number of containers, operation type (import/export), operation time (start/end), Date and time of arrival, berthing, and departure from port and berthing position. Actual data for one week operation from 3/3/2014 to 9/3/2014 which required for our proposed model were selected. A simulation model was developed and verified and validated using the terminal operational parameters.

**Experimental setting**

1. All resources from the same type have the same specifications.
2. A vessel is allocated to the berth depending on First Come First Serve rule.

**EXPERMINTAL RESULTS**

The input data for simulation models are based on real data at the El-Dekheilla port Container Terminal for one week from 3/3/2014 to 9/3/2014. This involved 8 ship arrivals, ship length (meter), start of operation time, end of operation time, no of imported (unloaded)/exported (loaded) containers and actual service time (minutes) for each ship according to Table 1.

Table 1: container terminal Actual operation

| Ship No | ship length(m) | operation time (Start) | operation time (End) | container (imp/exp) | No of containers | | actual service time (minutes) |
|---|---|---|---|---|---|---|---|
| | | | | | 20 ft | 40 ft | |
| 1 | 287 | 3/3/14 10:45 PM | 5/3/14 2:00 PM | Import | 442 | 236 | 2355 |
| | | | | Export | 365 | 567 | |
| 2 | 296 | 4/3/14 3:00 AM | 6/3/14 1:45 AM | Import | 539 | 464 | 2805 |
| | | | | Export | 101 | 452 | |
| 3 | 137 | 5/3/14 12:30 PM | 5/3/14 6:30 PM | Import | 6 | 13 | 360 |
| | | | | Export | 0 | 108 | |
| 4 | 183 | 5/3/14 6:15 PM | 6/3/14 8:00 AM | Import | 45 | 312 | 825 |
| | | | | Export | 160 | 218 | |
| 5 | 197 | 6/3/14 6:00 AM | 6/3/14 1:00 PM | Import | 44 | 345 | 420 |
| | | | | Export | 0 | 0 | |
| 6 | 101 | 8/3/14 9:15 AM | 8/3/14 1:45 PM | Import | 36 | 21 | 270 |
| | | | | Export | 6 | 21 | |
| 7 | 157 | 9/3/14 4:00 AM | 9/3/14 9:30 PM | Import | 22 | 88 | 1050 |
| | | | | Export | 219 | 231 | |
| 8 | 171 | 9/3/14 8:00 PM | 10/3/14 9:00 AM | Import | 143 | 99 | 780 |
| | | | | Export | 80 | 0 | |
| | | | | | | | 8865 |

Implementation of the model was run on a Laptop with the following configurations: i3 CPU 2.4 GHZ, 4.0 GB RAM, Windows 7 using discrete event simulation software (Flexsim). A distinct advantage of Flexsim software over similar software like Arena is that it comes with flexsim CT, a library specifically designed for simulating container terminal operation.

Table 2 shows the results of our proposed simulation model on a real case study data of container terminal at El-Dekheilla port operation for one week from 3/3/2014 to 9/3/2014. The results shows that our proposed simulation model achieved a total berthing time (service time) = 4322, where the actual service time of the collected data=8865 (see table 1).  Figure.3 illustrates a graph comparison between both actual and proposed model operation time (service time) versus TEU. From the figure it is obvious that our proposed simulation model enhanced the service time and achieved better results than the methods that container terminal planners use.

Table 2: A proposed model results

| Ship No | ship length(m) | operation time (Start) | operation time (End) | container (imp/exp) | No of containers 20 ft | No of containers 40 ft | Proposed model service time (minutes) |
|---|---|---|---|---|---|---|---|
| 1 | 287 | 3/3/14 10:45 PM | 4/3/14 6:17 PM | Import | 442 | 236 | 1172 |
|  |  |  |  | Export | 365 | 567 |  |
| 2 | 296 | 4/3/14 3:00 AM | 4/3/14 9:45 PM | Import | 539 | 464 | 1125 |
|  |  |  |  | Export | 101 | 452 |  |
| 3 | 137 | 5/3/14 12:30 PM | 5/3/14 4:25 PM | Import | 6 | 13 | 235 |
|  |  |  |  | Export | 0 | 108 |  |
| 4 | 183 | 5/3/14 6:15 PM | 6/3/14 4:25 AM | Import | 45 | 312 | 610 |
|  |  |  |  | Export | 160 | 218 |  |
| 5 | 197 | 6/3/14 6:00 AM | 6/3/14 12:16 PM | Import | 44 | 345 | 376 |
|  |  |  |  | Export | 0 | 0 |  |
| 6 | 101 | 8/3/14 9:15 AM | 8/3/14 11:15 AM | Import | 36 | 21 | 120 |
|  |  |  |  | Export | 6 | 21 |  |
| 7 | 157 | 9/3/14 4:00 AM | 9/3/14 9:04 AM | Import | 22 | 88 | 304 |
|  |  |  |  | Export | 219 | 231 |  |
| 8 | 171 | 9/3/14 8:00 PM | 10/3/14 2:20 AM | Import | 143 | 99 | 380 |
|  |  |  |  | Export | 80 | 0 |  |
|  |  |  |  |  |  |  | 4322 |

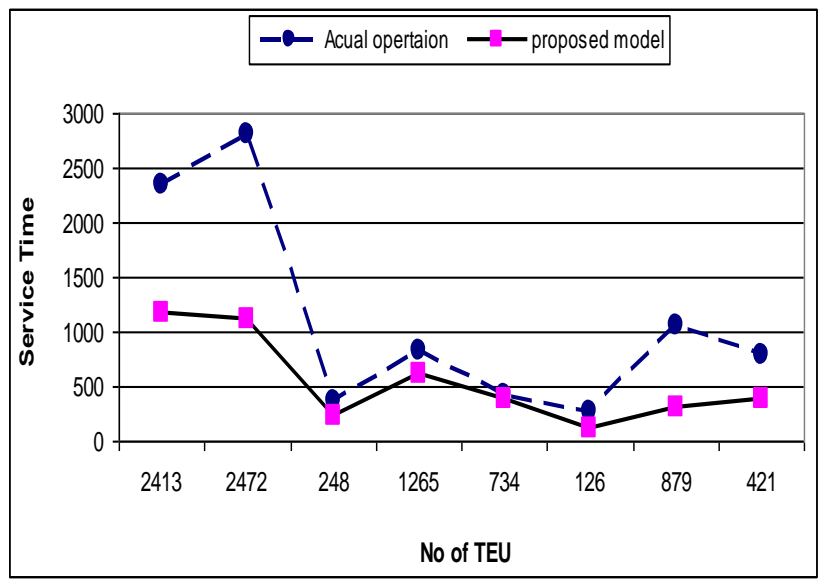

Figure.3 comparison between actual and proposed model operation time

## CONCLUSION AND FUTURE WORK

In this paper, we proposed a simulation model that can be used to optimize operations in Container Terminal using simulation methodology, minimize ship's turn-around time and improve the performance of the container terminal. The proposed approach is applied on a real data from Alexandria Container Terminal at El-Dekheilla port. Computational experiments were conducted to analyze the performance of container terminal operation. The results show that the proposed approach reduce the ship turnaround time in port where 51% reduction in ship service time (loading/unloading) in port is achieved.

In the future, analytical and empirical evaluations will be done for optimization of container handling system using simulation based optimization technique and develop a computational framework for enhancing the computational efficiency of the proposed solution technique.

32. Alexandria Container terminal: www.alexcont.com [accessed 15/5/2014]

**AUTHOR'S PROFILE**

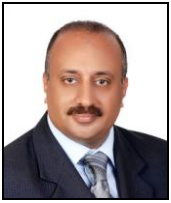

**Gamal Abd El-Nasser A. Said**: He received his M.Sc. (2012) ) in computer science from College of Computing & Information Technology, Arab Academy for Science and Technology and Maritime Transport (AASTMT), Egypt and B.Sc (1990) from Faculty of Electronic Engineering, Menofia University, Egypt. His work experience as a Researcher, Maritime Researches & Consultancies Center, Egypt. Computer Teacher, College of Technology Kingdom Of Saudi Arabia and Lecturer, Port Training Institute, (AASTMT), Egypt. Now he is Ph.D. student in computer science, Ain Shams University. His research areas include optimization, discrete-event simulation, and artificial intelligence.

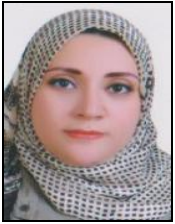

**Dr Abeer M. Mahmoud**: She received her Ph.D. (2010) in Computer science from Niigata University, Japan, her M.Sc (2004) B.Sc. (2000) in computer science from Ain Shams University, Egypt. Her work experience is as a lecturer assistant and assistant professor, faculty, of computer and information sciences, Ain. Shams University. Her research areas include artificial intelligence medical data mining, machine learning, and robotic simulation systems.

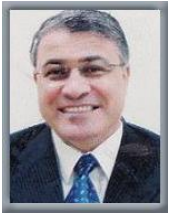

**Professor El-Sayed M. El-Horbaty**: He received his Ph.D. in Computer science from London University, U.K., his M.Sc. (1978) and B.Sc (1974) in Mathematics From Ain Shams University, Egypt. His work experience includes 39 years as an in Egypt (Ain Shams University), Qatar (Qatar University) and Emirates (Emirates University, Ajman University and ADU University). He Worked as Deputy Dean of the faculty of IT, Ajman University (2002-2008). He is working as a Vice Dean of the faculty of Computer & Information Sciences, Ain Shams University (2010-Now). Prof. El-Horbaty is current areas of research are parallel algorithms, combinatorial optimization, image processing. His work appeared in journals such as Parallel Computing, International journal of Computers and Applications (IJCA), Applied Mathematics and Computation, and International Review on Computers and software.